1    # Interaction of scanning probes with semiconductor nanocrystals;

2    ## Physical mechanism and basis for near field optical imaging

3    Yuval Ebenstein , Eyal Yoskovitz, Ronny Costi, Asaf Aharoni, Uri Banin[*]

4    Institute of Chemistry, the Farkas Center for Light Induced Processes, and the Harvey M.

5    Kreuger Family Center for Nanoscience and Nanotechnology, The Hebrew University of

6    Jerusalem, Jerusalem 91904, Israel

1    **Abstract**

2    We investigate the modification of photoluminescence (PL) from single semiconductor

3    nanocrystal quantum dots (NCs) in proximity of metal and semiconducting Atomic Force

4    Microscope (AFM) tips. The presence of the tip alters the radiative decay rate of an emitter via

5    interference and opens efficient non radiative decay channels via energy transfer to the tip

6    material. These effects cause quenching (or enhancement) of the emitter's PL intensity, as a

7    function of its distance from the interacting tip. We take advantage of this highly distance

8    dependent effect to realize a contrast mechanism for high resolution optical imaging. AFM tips

9    are optimized as energy acceptors by chemical functionalization with InAs NCs to achieve

10   optical resolution down to 30 nm. The presented experimental scheme offers high resolution

11   optical information while maintaining the benefits of traditional AFM imaging. We directly

12   measure the PL intensity of single NCs as a function of the tip distance. Our results are in good

13   agreement to calculation made by a classical theoretical model describing an oscillating dipole

14   interacting with a planar mirror.

16   **Introduction**

17   The presence of a dielectric tip causes strong modification of photoluminescence (PL) properties

18   from nearby fluorophores[1, 2]. This effect has been exploited to generate contrast for apertureless

19   near-field scanning microscopy (A-NSOM) schemes [3-5].   Near-field microscopy provides

20   enhanced resolution beyond the diffraction limit of far-field microscopy.   In conventional

21   NSOM[6], light is driven through a sub-wavelength aperture and practically, the resolution is

22   limited by the size of the aperture and the optical skin depth of the metal coating, typically on the

23   order of 100 nm[7].   In apertureless near field microscopy, the resolution is obtained by

1    perturbation of the optical field by a nm sized tip in the proximity of the sample and resolution

2    down to 1nm was demonstrated[8].  Previous works measuring PL with A-NSOM reported a

3    variety of optical phenomena such as PL enhancement and PL quenching. Various lateral PL

4    distribution patterns were observed for seemingly similar experiments in which the PL of

5    emitting specimens was measured with respect to a scanning tip position. Some of these effects

6    have been explained by differences in experimental conditions such as the mode of excitation, tip

7    treatment, and sample properties[9]. Recently, finite difference time domain simulations of such

8    dipole-tip systems demonstrated that probe geometry has a substantial effect on the distance

9    dependence of the interaction, shedding light on some of the observed PL patterns[10]. Here we

10   investigate systematically the effect of PL quenching by AFM tips of different materials

11   including tips coated by semiconductor nanocrystals (NCs)[11]. We present an experimental

12   scheme enabling simultaneous acquisition of AFM and apertureless near field optical images

13   while recording the distance dependent behavior of the tip-sample system.  The availability of

14   this data for each point of the image provides insight on the nature of the tip induced optical

15   interaction with clear relevance to high-resolution microscopy schemes.

16      With the rapid progress of nanoscience and related technologies, efforts are made to enhance

17   the available tool box for the study of nano systems and larger complex systems where nanoscale

18   detail is required. The combination of AFM and high resolution optical data measured

19   simultaneously presents a powerful approach in this context[12, 13].  Tapping mode AFM is the

20   technique of choice for the study of weakly bound nanostructures and soft biological samples, as

21   it does not inflict lateral forces on the sample.  In this case, however, the optical interaction is

22   averaged out due to the tip-sample distance modulations and specific experimental schemes are

23   required to extract meaningful optical information[14, 15]. Such a scheme is developed in this study

1   allowing extraction of distance dependent optical data, using distance correlated single photon

2   counting (DCSPC). We applied the method to the investigation of semiconductor NCs which

3   posses both measurable topography and fluorescence tunable by size, composition and shape.  In

4   previous work we already pointed out the potential use of NC functionalized tips for a

5   microscopy scheme based on fluorescence resonant-energy transfer (FRET)[11]. Such tips are used

6   here and their optical interaction with the sample is characterized.

7        The obtained distance-dependent optical data provides insight to the interaction mechanisms

8   of the fluorophore with the tip material and can be understood within the framework of

9   fluorophores in front of a dielectric mirror.  The radiative and nonradiative decay rates of an

10  excited dipole fluorescing proximal to a metal film have been extensively investigated, both

11  theoretically and experimentally[16-20]. The best agreement between experiment and theory was

12  achieved by Chance, Prock and Silbey (CPS) who calculated the total energy flux through

13  infinite planes above and below the dipole[17]. This method yields separate expressions for the

14  radiative and nonradiative decay rates thus enabling the formulation of expressions both for the

15  lifetime and for the luminescence efficiency of the dipole as a function of its distance from an

16  interface. The lifetime of emitting molecules and NC ensembles in front of a dielectric interface

17  has previously been reported[16, 17, 19-22].

18       More recently, the lifetime and intensity of single molecules and NCs in proximity to an AFM

19  tip have been investigated[5, 14, 15, 23, 24]. These measurements have shown both quenching and

20  enhancement of the fluorescence intensity depending on experimental conditions.  We present

21  measurements of the distance dependent PL intensity from single NCs, and compare our results

22  with the theoretical predictions for the quantum efficiency of a dipole in front of a mirror using

23  the CPS theory. The strong modulation of PL intensity in the presence of the tip is exploited for

1    optical imaging based on PL quenching and provides a basis for future high-resolution

2    microscopy schemes.

4    **Experimental Details**

**Fig 1.**

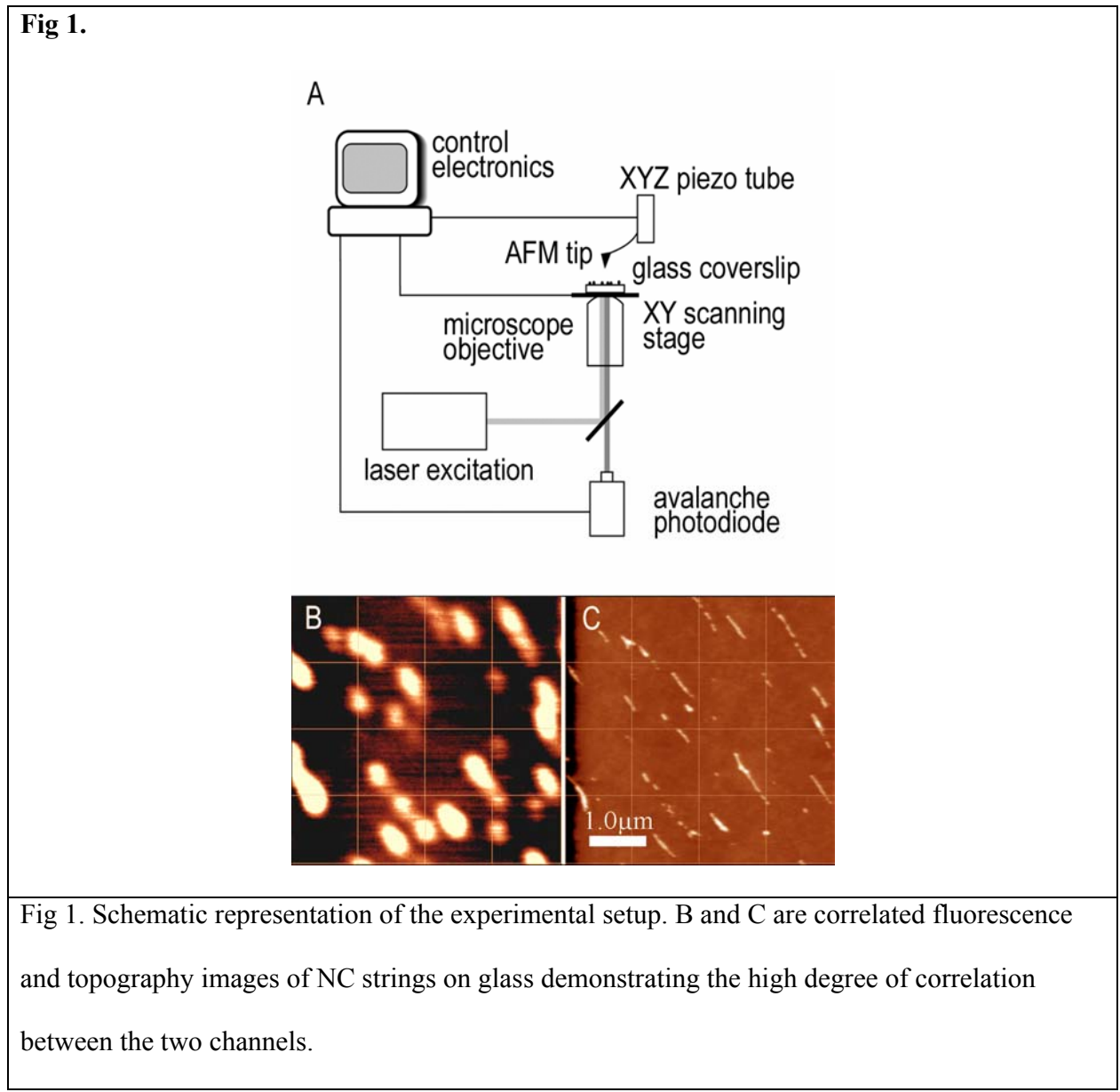

Fig 1. Schematic representation of the experimental setup. B and C are correlated fluorescence

and topography images of NC strings on glass demonstrating the high degree of correlation between the two channels.

1    The experimental setup (Fig 1) comprises of an AFM head (Veeco, Bioscope) mounted on top of

2    an inverted optical microscope (Zeiss, Axiovert 100), equipped with a piezoelectric scanning

3    stage (Nanonics, Flatscan) enabling simultaneous acquisition of AFM and optical data[12]. An

4    example for a correlated PL-topography measurement of NCs on glass is shown in Fig1B. As an

5    excitation source, the 514 nm line of an Ar-ion laser is passed through a single-mode fiber for

6    spatial filtering, and tuned to circular polarization. The beam is collimated onto the back aperture

7    of an oil immersion, high numerical aperture (NA) objective (Zeiss, X100, NA 1.4), and focused

8    to a diffraction limited spot on the sample. Fluorescence is collected by the same objective in an

9    epi-illumination configuration, and filtered from the excitation light by passing through a

10   dichroic mirror and a 514nm Raman-edge filter. Collected fluorescence is focused on to the

11   active area of an avalanche photodiode (APD) (Perkin-Elmer, SPCM-ARQ 13). In a typical

12   measurement, a dilute solution of colloidal NCs (CdSe/ZnS, Core/Shell dots and rods) is spin

13   cast on a clean glass coverslip. The sample is raster scanned across the excitation spot until NC

14   fluorescence is detected. The scan is then stopped with the NC is parked within the excitation

15   spot. At this point, the AFM tip is scanned over the particle, and fluorescence photons are

16   collected pixel by pixel in correlation with the tip lateral position. As the tip approaches the NC,

17   the fluorescence count rate is modified and optical contrast is created in the collected photon

18   distribution map. (A schematic animation is presented in supplementary flash movie 1).

19       Earlier works[4, 23] performed such experiments with a fixed tip-sample distance achieving

20   optical contrast for single molecules. We utilize tapping mode AFM (TAFM) combined with

21   time correlated single photon counting in order to obtain the distance dependence of the optical

22   process during imaging where the counter computer card is triggered by the tip oscillation. This

23   distance correlated single photon counting scheme (DCSPC) allows us to optimize the optical

Interaction of scanning probes with semiconductor nanocrystals

1    contrast by choosing optical data acquired at specific tip-sample distance regions as indicated by

2    the distance dependence measurements.

3        Three types of AFM tips were investigated including bare Si tips, platinum coated tips, and

4    NC functionalized tips[11].  NC functionalized tips were prepared by coating the tips first with a

5    layer of MPTMS (mercaptopropyl-trimethoxysilane) linker molecules followed by immersion in

6    a nanocrystal solution.

7    **Results and Discussion**

8        Two sets of experiments are presented; the first conducted in contact mode AFM and the

9    second in our combined TAFM-DCSPC scheme. By monitoring the optical PL images of single

10   emitting NCs we compare the quenching properties of the three types of AFM tips mentioned

11   above.

**Fig 2.**

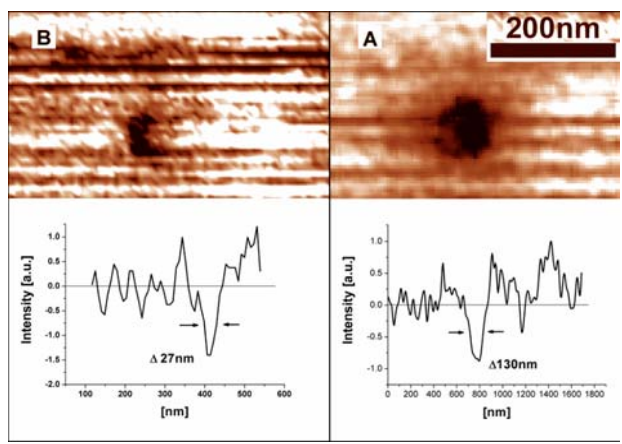

Fig 2. Quenching spots and corresponding fluorescence intensity cross sections of single NCs

interacting with: A- a Platinum coated tip (130nm FWHM). and B- an InAs functionalized tip

(27nm FWHM); Bottom shows corresponding cross sections through the quenched spots.

Interaction of scanning probes with semiconductor nanocrystals

1    Fig.2 shows typical optical images of single CdSe/ZnS NCs acquired in contact mode. In this

2    mode, the tip apex maintains physical contact with the sample surface thus maximizing the

3    optical interaction. Since the tip induces lateral forces during scanning, the NCs are embedded in

4    a ~10nm polymethylmetacrylate (PMMA) film to prevent sweeping of the particles to the scan

5    borders[25]. Fig 2A was recorded using a commercially available Platinum coated tip

6    (Mikromasch, CSC12). A distinct dark spot is apparent in the image corresponding to the

7    position of the embedded NC. The size of these dark spots is typically on the order of 140nm

8    FWHM, well below the diffraction limit and the size of our excitation spot which is in the order

9    of 350nm FWHM. The image in Fig 2B was recorded using a silicon tip (Mikromasch, CSC12)

10   coated with a sub-monolayer of InAs NCs via chemical functionalization as described above[10].

11   These NCs were chosen since they are assumed to act as excellent energy acceptors for the

12   CdSe/ZnS NCs in the sample. They posses a strong absorption cross-section at the region of

13   interest, and emit weakly at ~1150nm, outside our detection window (emission and absorption

14   spectra for this NC configuration is shown in supplementary Fig 1.). Using such functionalized

15   probes resulted in quenched-fluorescence spots with diameters on the order of 30nm FWHM,

16   considerably smaller than those recorded with the metal coated tips. We note that experiments

17   executed with bare silicon tips did not result in tip related optical contrast.

18   There are two possible explanations for the consistent difference in resolution between the

19   metal coated and NC functionalized tips; first, the physical dimension of the tip's interaction

20   region is larger in the case of the metal coated tip. Scanning electron microscope images of the

21   two tip kinds show that there is indeed a difference in tip diameter between the two tips, as can

22   be expected from the tip preparation process. While both tips are based on similar Silicon tip

23   template, platinum coated tips are significantly larger because of the evaporated metal film that

1   is a few tens of nm thick. The other tip type is chemically coated with a sub-monolayer of NCs

2   with a diameter of ~5nm, keeping the sharpness of the original tip.

3      A second explanation for the improved resolution is related to the details of energy transfer

4   processes between the NC fluorophore on the surface and the quenching tips.  As discussed

5   below, calculations of the characteristic energy transfer distance for platinum versus InAs tip

6   material could not explain the 5 fold improvement in resolution for the InAs functionalized tips.

7   Furthermore, using the CPS model to describe both systems (solid and dashed lines in Fig 5B)

8   we find that energy transfer occurs actually at shorter distance for the platinum case which is in

9   contradiction with our observations. A possible explanation for this discrepancy is that since the

10  functionalized tip is composed of isolated NCs, we cannot assume it is well represented by the

11  interaction of a chromophore donor with a planar acceptor surface described by the CPS theory

12  (2-dimensional picture).  A more appropriate picture for the NC functionalized tips may be of an

13  energy transfer process between two point dipoles. While the energy transfer efficiency in the 2-

14  dimensional case has $1/d^3$ dependence on distance $d$, the dipole-dipole case takes the familiar

15  Foerster form with $1/d^6$ distance dependence.  This would lead to a steeper quenching profile and

16  better resolution.

17     As a consequence of the polymer embedding needed for contact mode operation of the AFM,

18  topography and other high resolution information is not accessible in the former pictures.

19  Furthermore, since the tip remains in a fixed distance from the emitting dipole, the detailed

20  distance dependence of the interaction may not be studied. As mentioned earlier, tapping mode

21  AFM operation is desirable to provide topography as well as distance dependent interactions.  To

22  achieve this we combine the AFM measurement with time correlated single photon counting

23  (TCSPC). Briefly, an electrical trigger pulse is generated with each oscillation of the AFM tip

1    and fed as a synchronization signal to a TCSPC computer card (Picoquant, Time Harp 200). The

2    arrival time of each photon collected by the APD is recorded relative to the tip signal with 1nsec

3    resolution and a histogram of photon arrival times is generated for each pixel in the image. Since

4    the frequency and amplitude of oscillation are known, the histograms are easily converted to

5    approach curves representing photon-count Vs. tip-sample distance yielding a distance correlated

6    single photon counting scheme (DCSPC).

**Fig 3.**

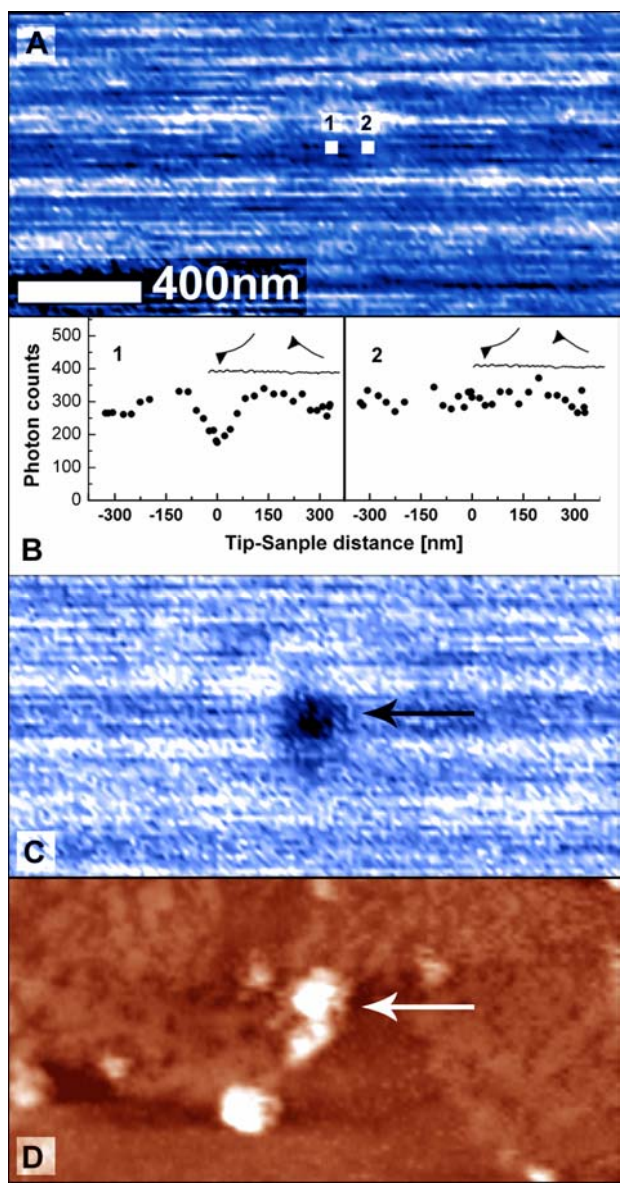

Fig 3. Signal to noise enhancement through generation of a distance dependent fluorescence image. A-Fluorescence image comprised of all the collected photons; B-Photon counts as a function of tip-sample distance while tapping; 1-taken from an interaction region and showing distinct fluorescence quenching. 2-taken from a nearby area and showing no distance dependent

Interaction of scanning probes with semiconductor nanocrystals interaction with the sample; C-Fluorescence image comprised of photons collected from the bottom 10nm of the tip oscillation; D-Simultaneously acquired AFM topography image of NCs on glass. Arrow marks the quenched particle.

2      Fig 3. shows a representative optical image using the DCSPC scheme. In the upper frame

3    (Fig. 3A), the far-field confocal image containing all the photons recorded is presented. The NC

4    jumps between high and low count rate states resulting in the visible streaks in the recorded

5    optical image. This "blinking" phenomenon is typical to single NC emission[26]. The oscillation

6    amplitude of the tip in this case is 350nm and no optical contrast can be resolved in the image,

7    due to integration of the response at different tip-sample distances which smears the effects

8    recorded at very short distances. In Fig. 3B, two approach curves of photon counts vs. tip

9    position are presented corresponding to the two pixel positions designated (1) and (2) in the

10   optical image. Curve (2) shows no distance dependent characteristics while curve (1) clearly

11   shows a systematic dependence of the photon counts in respect to the tip-sample distance. The

12   emission intensity is dramatically decreased as the tip approaches the NC. Since each photon is

13   recorded with its respective tip position, it is now possible to generate an image composed only

14   of the photons correlated to a desired range of tip-sample distances.  The image in Fig. 3C is

15   composed only of the photons arriving while the tip was at the lowest 10nm of its oscillation. A

16   clear region of quenched fluorescence is revealed, demonstrating the enhancement of the signal

17   to noise ratio relative to the unprocessed image. The simultaneously acquired topography image

18   of NCs on glass is presented in Fig. 3D. An arrow marks the particle under investigation.

19   Clearly, the acquisition of topography and other AFM data in conjunction with the optical

20   information enables the assignment of optical data to specific regions of interest indicated by

1    AFM. While the far-field optical data gives chemical and other information common for

2    fluorescence microscopy, the distance dependent measurement offers the possibility to localize

3    the emitting specie and to gain additional information from its distance dependent behavior.

**Fig 4.**

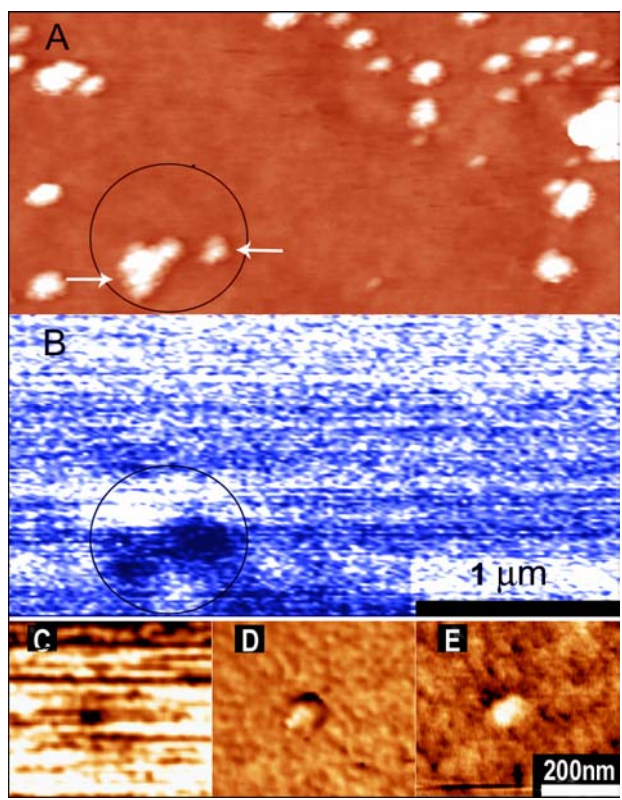

Fig 4. NCs on glass: Topography (A) and distance dependent fluorescence (B) images acquired

with a platinum coated tip. Two quenching spots optically distinguish between emitters in the far field excitation region indicated by circles.

C,D and E are Fluorescence, phase and topography images respectively, of a single NC acquired with an InAs functionalized tip. The quenching spot in C, has a diameter of ~60nm FWHM.

1    Two examples for the correlation between AFM and optical images are shown in Fig 4. In

2    Figs. 4A and 4B the images were acquired by a Platinum coated tip. The topography image

3    shows two NC aggregates within the excitation spot indicated by the circle. In a far-field image,

4    the two aggregates would appear as a single fluorescence spot. Nevertheless,  the two aggregates

5    are clearly distinguishable in the quenching induced optical image; moreover, the left quenching

6    spot correlates to a specific region in the bigger of the two aggregates indicating that it is

7    composed only in part of emitting particles (marked with an arrow). Images obtained with a tip

8    functionalized with InAs NCs, are presented in the lower panels (Figs. 4C,D & E corresponding

9    to optical, phase and height images, respectively), significantly improved optical resolution of

10   60nm FWHM is achieved for a single NC (Fig. 4C). This resolution is similar to that of the

11   correlated phase (Fig. 4D) and topography (Fig. 4E) scans.  This improved resolution for InAs

12   tips is consistent with our results in contact mode for the two tip types.

13   The images demonstrate the applicability of the method for optically resolving the location of

14   fluorescent markers with high precision. This is especially attractive for biological samples such

15   as DNA molecules and cells where physical structure may be determined by various AFM

16   techniques while the optical data could provide also localized chemical information through the

17   quenching of markers bound to the studied specimen.

18   An additional important feature of the experimental setup we developed is the availability of

19   both AFM and distance dependent optical data for every pixel in the image.  This enables the

20   detailed analysis of approach curves correlated to specific locations in the topographic image.

21   Specifically, we employ this to study the modification of PL intensity from single NCs

22   interacting with the AFM tip. We focus on the investigation of energy transfer and quenching of

23   PL by the approaching tip. We eliminate most enhancement effects by using in-plane polarized

1    excitation thus minimizing the coupling and concentration of the optical field at the tip apex.

2    Nevertheless, due to the high N.A of our microscope objective, a polarization component

3    perpendicular to the sample plane exists.

4        The measurements are accompanied by theoretical analysis within the CPS framework.  We

5    model our system as a flat dielectric mirror (tip) approaching a single dipole (NC), fixed in

6    vacuum (air) and parallel to the mirror. The total energy flux out of the upper and lower planes is

7    calculated using the complex pointing vector and integrating its normal component over the

8    plane[27]. Dividing the energy flux by the dipole energy gives the rate constant associated with

9    energy loss through the plane. With these rate constants we can now calculate the lifetime and

10   quantum efficiency of the system for any given distance[28]. Using the CPS model, we calculate an

11   apparent quantum efficiency (aQE) which is directly proportional to the PL intensity measured

12   by integrating over all photons emitted to the half space below the dipole. We justify our choice

13   of the parallel dipole model by the following: 1) the calculated aQE for a parallel dipole is much

14   larger than that for the perpendicular one making our measurement more sensitive to parallel

15   dipole emission. And 2), the 2D degenerate nature of the NC emission dipole[29, 30] ensures that

16   there is always a parallel emission component to be detected.

**Fig 5.**

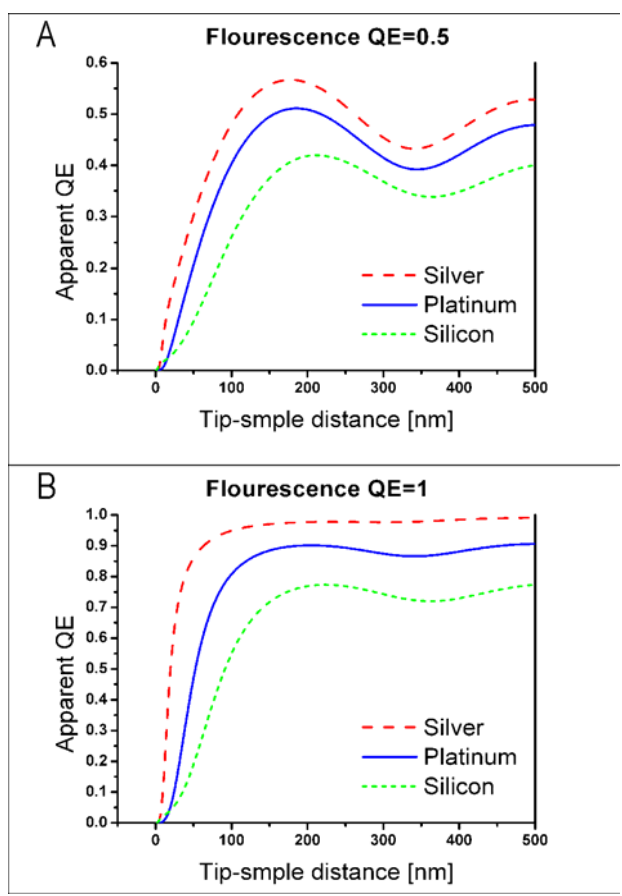

Fig 5. Calculated apparent quantum efficiency for a dipolar emitter with an unperturbed quantum efficiency of 0.5 (A) and 1 (B). The dipole is approached by a planar mirror of 3 different materials: Silver (dashed line), Platinum (solid line) and Silicon (doted line).

2      Figure 5 shows the results of such calculation for various tip materials presenting the aQE

3   versus tip-sample distance.  At short ranges, most relevant to our present application, there is a

4   steep reduction in aQE due to the efficient energy transfer to the tip.  At longer distances, some

5   oscillatory behavior is predicted due to interference of the dipole with its image dipole in the

6   dielectric material.  Among the three calculated materials, silver shows a very steep response that

1   could potentially translate into higher resolution.  Practically Ag is a problematic coating

2   material for the tip due to its tendency for oxidation in ambient conditions.  Another potential

3   issue with silver is the possibility of enhanced emission via plasmonic effects which are not

4   treated here. An additional variable influencing the calculation is the unperturbed QE of the

5   emitting dipole as can be seen comparing the distance dependence for QE=0.5 (Fig. 5A) with

6   QE=1 (Fig. 5B). The effect of varying the dipole QE is manifested in the slope of the final

7   150nm of the tip-sample approach curve. As nonradiative energy transfer becomes the dominant

8   interaction mechanism, the slope becomes steeper for higher QEs.

**Fig 6.**

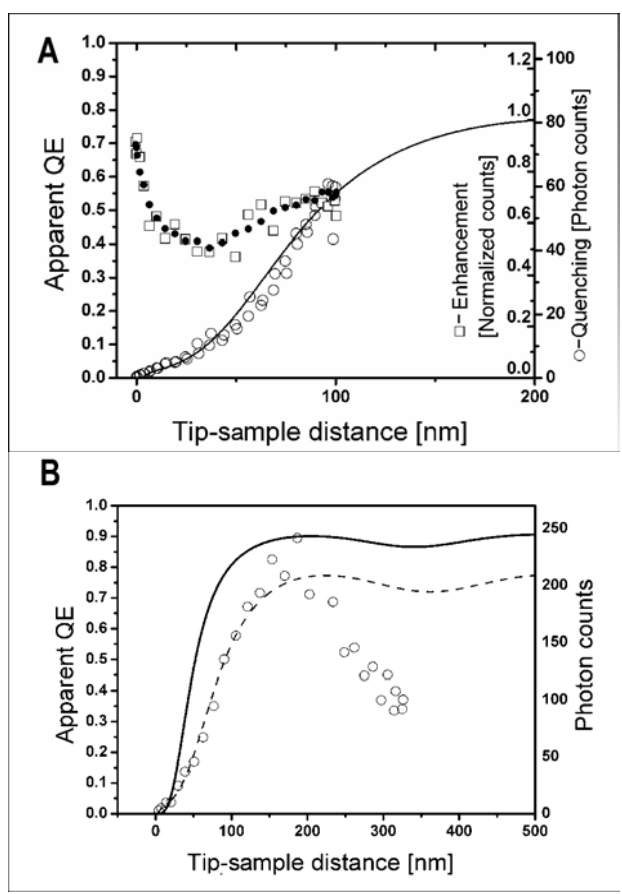

Fig 6. A- Measured DCSPC traces with silicon tips. Open circles: measured photon counts versus distance showing quenching. Open squares: normalized average of 5 measurements showing enhancement. Black dots: 4 point averaging of the enhancement data. In solid lines we plot the calculated apparent QE of a Silicon mirror approaching an emitting dipole parallel to the surface. Good agreement to the measure data is observed.

The dipole QE is assumed to be 1.

B- Measured photon counts of an NC as a function of the distance from a Platinum coated tip (open circles). We plot the calculated apparent QE of a Platinum (solid line) and Silicon (dashed line) mirror approaching an emitting dipole parallel to the surface. A better agreement to the silicon calculation up to 150nm is observed. Beyond this distance the reduction observed experimentally is not described by the model. (See text).

2    In Fig 6 we present typical approach curves measured with silicon and platinum-coated tips

3    approaching a NC with a 605nm emission peak. All curves are measured during normal tapping-

4    AFM operation with an average of 5-10 ms per pixel to demonstrate the possibility for using this

5    approach in practical imaging schemes with a frame time of a few minutes. For the approach

6    curves presented here we used a 2*2 pixel binning, i.e. 20-32 ms integration time per curve. By

7    measuring the scattered excitation intensity as a function of tip distance we derived a correction

8    factor for the raw PL intensity data. This distance dependent factor was used to correct the

9    measured PL data for the change in excitation power due to backscattering of excitation photons

10   from the tip. In addition we plot the calculated apparent QE of an emitting dipole oriented

11   parallel to the surface as it is approached by a planar silicon or platinum mirror. The theoretical

12   curves use optical constants for the tips materials (Si, Pt or InAs) taken from the literature[31, 32].

13   The measured QE in solution for the NC sample was 0.5 and QE studies of single NCs support in

14   this range a QE value of 1 for the emitting particles[12, 33].  We therefore use a value of 1 for the

15   unperturbed QE in the calculations as well.

16   In fig 6A we present results for a silicon tip. Experimental and calculated DCSPC approach

17   curves for InAs functionalized tips were found to be almost identical to those of bare silicon and

18   therefore are not presented. Although the CPS model is only a first approximation for the NC –

19   tip system, experimental data (open circles) follows reasonably well the theoretical curve (solid

20   line) for the 100nm region measured. These efficient quenching properties are in contradiction to

1    our observation of poor contrast in the silicon quenching images in comparison with the

2    functionalized or metalized tips. Generating approach curves from nearby pixels in the Si case

3    (open squares) reveals an additional effect of fluorescence enhancement by the proximity of the

4    silicon tips. This enhancement effect for silicon tips has been previously reported[9, 34] and

5    exploited for high resolution optical imaging[11]. We believe that this field enhancement by the

6    sharp Si tip is caused by out of plane polarization components in the tightly focused excitation

7    spot. The coexistence of both quenching and enhancement result in low optical contrast for

8    imaging with silicon tips, unlike platinum and NC coated tips which give good contrast and

9    present no enhancement effects in our experimental conditions.

10    It is possible that when the silicon tips are functionalized with InAs NCs, the latter serve to

11    suppress the enhanced field at the tip apex in addition to their role as efficient energy acceptors

12    for the studied sample PL. This twofold activity gives rise to the high quenching contrast visible

13    in the images acquired with functionalized tips as apposed to bare silicon tips. In fig 6B, an

14    average of three approach curves from different particles measured with a platinum coated tip

15    (open circles) is plotted together with the calculated curves for platinum (solid line) and silicon

16    (dashed line) mirrors. Here too, experimental data follows the silicon mirror model for the first

17    150nm. The reason for the silicon like behavior of the platinum coated tip may lie in the fact that

18    in our model we regard the tip as bulk platinum, neglecting the fact that it is composed of silicon

19    coated by ~30 nm of platinum with a titanium binding layer. Given the optical skin depth of

20    platinum in this spectral region (few tens of nm) [35-37] it is possible that the silicon tip still has a

21    significant effect on the quenching process. Furthermore optical parameters for thin metal films

22    depend strongly on film preparation and should be regarded with caution[30].

1    Beyond 150 nm, we observe experimentally a decrease in count rates that deviates from the

2    calculated behavior. To address this drop in photon count rate at longer tip-sample distances we

3    note another difference between our approximate model and the real system. It has been shown

4    both experimentally and theoretically[21] that the emission pattern, i.e. the spatial distribution of

5    emitted photons from a dipolar emitter, is also strongly dependent on its distance from the

6    interface. While our model calculates the contribution of all photons emitted to the half space

7    below it, in the experiment we are restricted to collection from a finite solid angle determined by

8    our microscope objective. A recent experiment by Buchler et al[38], featuring a scanning micro

9    mirror – dye system and using a model containing variation in emission pattern, has indeed

10   showed strong intensity modulations even for a high QE emitter related to this effect.

12   **Conclusion**

13   We have demonstrated how tapping mode AFM and fluorescence microscopy with distance

14   correlated single photon counting may be combined to generate a wealth of information in a

15   single correlated measurement. The combined tool presented here offers the detailed structural

16   characterization of AFM with the chemical information available by the optical microscope and

17   the use of fluorescent tagging. In addition, the presence of the AFM tip enables both high

18   precision optical localization of emitting species and a detailed study of distance dependent

19   optical processes. Specifically, we have shown how energy transfer from emitting NCs to

20   platinum-coated or NC functionalized AFM tips results in quenching of PL in a distance

21   dependent manner. This process is appropriate as a contrast mechanism for high-resolution

22   appertureless near field imaging. Since florescence photons are collected and registered as a

23   function of tip position, specific slices of the optical image may be generated, resulting in high

1    optical contrast. In addition to enhancing the signal to noise ratio of the image, this data

2    registration may be used for the 3d reconstruction of optical fields. The resulting tip-sample

3    approach curves were compared to calculations based on the CPS model and were found to be in

4    reasonable agreement with the proposed model. Deeper theoretical understanding of these

5    processes will enable extracting yet additional information such as dipole orientation and QE; all

6    in a single measurement and on time scales common in practical scanning probe microscopy.

8    **Acknowledgments**

9    This research was funded in part by the Israel Science Foundation (grant #924/04 ) and by the

10   DIP (German-Israel Program).

13   **Supplementary Fig 1.**

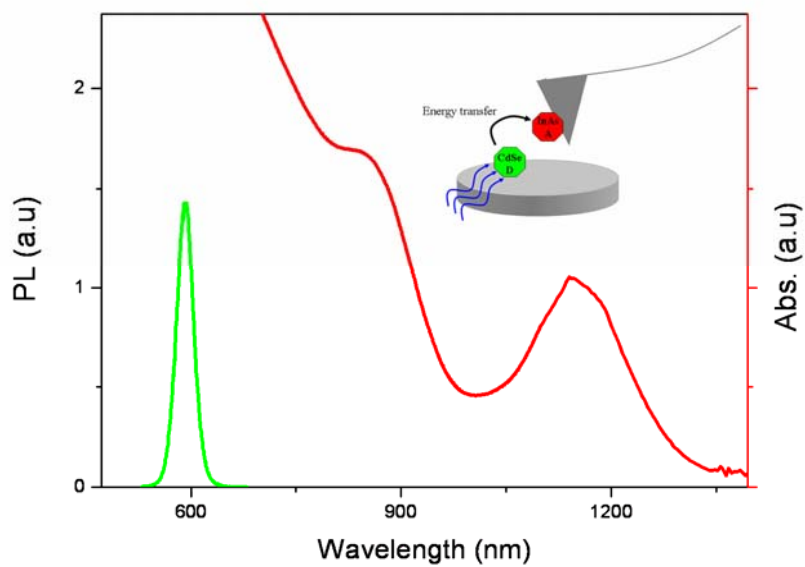

14   Interaction of scanning probes with semiconductor nanocrystals

Absorption spectrum of the InAs NCs used to functionalize the Silicon AFM tip (red) and the emission spectrum of the studied CdSe NCs on the substrate (green). A schematic of the experimental scheme is drawn in the inset.

Interaction of scanning probes with semiconductor nanocrystals

Interaction of scanning probes with semiconductor nanocrystals